\documentclass[aps,showpacs,preprintnumbers,amsmath,amssymb,
eqsecnum, twocolumn, tightenlines
]{revtex4}

\usepackage{graphicx}

\newcommand{\be}{\begin{eqnarray}}
\newcommand{\ee}{\end{eqnarray}}
\newcommand{\no}{\nonumber \\}
 \newcommand{\gsim}{\mathrel{\hbox{\rlap{\lower.55ex \hbox {$\sim$}}
                   \kern-.3em \raise.4ex \hbox{$>$}}}}
\newcommand{\lsim}{\mathrel{\hbox{\rlap{\lower.55ex \hbox {$\sim$}}
                   \kern-.3em \raise.4ex \hbox{$<$}}}}

\newcommand{\ba}{\begin{eqnarray}}
\newcommand{\ea}{\end{eqnarray}}

\def\bea{\be}
\def\eea{\ee}

\def\del{{\partial}}

\def\roughly#1{\mathrel{\raise.3ex\hbox{$#1$\kern-.75em%
\lower1ex\hbox{$\sim$}}}}
\def\lsim{\roughly<}
\def\gsim{\roughly>}

\def\({\left(}
\def\){\right)}
\def\[{\left[}
\def\]{\right]}

\def\la{{\Big<}}
\def\ra{{\Big>}}
\def\lsim{\mathrel{\rlap{\lower3pt\hbox{\hskip1pt$\sim$}}
     \raise1pt\hbox{$<$}}} 
\def\gsim{\mathrel{\rlap{\lower3pt\hbox{\hskip1pt$\sim$}}
     \raise1pt\hbox{$>$}}} 

\def\af{\alpha}
\def\bt{\beta}

\def\Lam{\Lambda}
\def\dlt{\delta}

\def\eps{\epsilon}
\def\omg{\omega}

\def\tz{\tilde{z}}
\def\lab{\label}
\def\le{\left}
\def\ri{\right}
\def\bea{\begin{eqnarray}}
\def\eea{\end{eqnarray}}

\begin{document}


\title{ Instabilities near  the QCD phase transition in the holographic models }

\author {  Umut G\"ursoy$^1$, Shu Lin$^2$ and   Edward Shuryak$^3$ }
\affiliation { $^1$ Institute for Theoretical Physics, Utrecht University, Leuvenlaan 4, 3584 CE Utrecht, The Netherlands}
\affiliation { $^2$RIKEN BNL Research Center, Brookhaven National Laboratory, Upton, NY 11973, USA}
\affiliation { $^3$ Department of Physics and Astronomy, State University of New York,
Stony Brook, NY 11794, USA}
\date{\today}

\begin{abstract}
The paper discusses phenomena close to the critical QCD temperature, using the holographic model.
One issue studied is the overcooled high-T phase, in which we calculate quasi normal sound modes. 
We do not find instabilities associated with other first order phase transitions, but nevertheless observe drastic changes in
sound propagation/dissipation. The rest of the paper considers a cluster of the high-T phase in the UV in coexistence with the low-T phase,
in a simplified ansatz in which the wall separating them is positioned only in the holographic coordinate. 
This allows to find the force on the wall and classical motion of the cluster. When classical motion is forbidden, we
evaluate tunneling probability through the remaining barrier.
\end{abstract}
\maketitle

\preprint{RBRC-1034, ITP-UU-13/24, SPIN-13/17 }



%









%

%

%





\section{Introduction}
\subsection{Motivation and overview}
    Production of a new phase of matter -- Quark-Gluon Plasma (QGP) is the main goal of ongoing experiments with Heavy Ion Collisions, currently performed
    at RHIC and LHC colliders.  During these collisions, the matter relatively rapidly equilibrates to QGP phase: but subsequent expansion leads to cooling.
    As a result, the temperature decreases with time, crossing back to the so called hadronic phase at temperature $T<T_c$. 
    In this paper we focus on the phenomena in the vicinity of the critical temperature $T_c$. 
    
More specifically, we study two distinct situations, to be referred to as (i) the ``beginning" and (ii) the ``end", of inhomogeneous
     matter distribution. 
     
    (i) The ``beginning" stage starts with a
    homogeneous high-$T$ phase which is overcooled and starts developing the inhomogeneous phase.      
    Certain instabilities which lead to the formation of spatially inhomogeneous
states are  well known in many fields of physics (for example, overcooled water vapor leads to formation of rain drops). General theory
of instabilities near many first order transitions is well developed. So one can be naturally interested if such phenomena
may also occur in a holographic setting.

   The growth of perturbations proceed to the so called 50-50 state, in which both phases occupy comparable fractions of the bulk:
   we don't attempt to discuss this stage.  Our interest will be 
   (ii) the near-end situation, in which the high-$T$ phase is represented  by a dilute gas of remaining clusters.
    As the expansion/cooling proceeds further, those have to disappear as well. 
   The exact fate of the clusters depends however on the parameters of the problem: they can either 
  become {\em unstable} and start collapsing due to classical equations of motion, or become  {\em metastable} 
   and decay via the thermal/quantum tunneling through the 
   remaining barrier.  Which route the system takes is a subject of the second part of this paper.

   From known examples one can infer that their decay
can be rather dramatic. One classic example is the Rayleigh bubble collapse,
capable of destroying hard steel of navy ship propellers. Another is sonoluminiscence phenomena (see e.g.\cite{Brenner:2002zz} for a review) concentrating the bubble 
in a volume of $\sim 10^{-6}$ of the original one and creating much higher temperatures (and light) from 
the room temperature water under the influence of just low-amplitude  low-frequency sound. 
This however happens only in well tuned conditions, as the results depend strongly  on bulk thermodynamical
quantities, as well as  on other essential ``details" such as 
the interphase surface tension, dissipative parameters
(viscosities), etc.

  Attempts to find observable consequences of the QCD critical behavior  
  as its temperature passes through the phase transition region $T\approx T_c\approx 170\, MeV$, from QGP to the hadronic phase, 
  have certain history. One of us 
 \cite{Shuryak:1997yj} had suggested to detect enhanced event-by-event critical fluctuations. Those  are expected to be enhanced 
near the hypothetical second-order QCD critical point  \cite{Stephanov:1998dy}: this idea had
motivated the  downward energy scan program at RHIC. While some changes in fluctuation pattern has been observed,
the program is not yet completed and the information about  location of the critical point (or even if it exists at all, in the domain covered by this program) remains 
inconclusive.
   
The existence of QGP clusters at the end of the expansion is  phenomenologically based on rapidity correlations between the
detected hadrons in heavy ion experiments.  
Recently  it has been proposed  \cite{Shuryak:2013uaa} 
search for signals of the sound emitted in the collapse of the QGP clusters.

    Depending on the rate of expansion, the instability can proceed differently. One scenario is that the free energy minimum
may turn at sufficient supercooling into a maximum. If this happens, one should be able to find a {\em classical instability} of the corresponding excitation mode. In section \ref{sec_modes} we will study the sound mode.
We  don't find density instability to be present, as the imaginary part of the quasi-normal modes (QNM) is
always negative, but we do observe rather drastic changes in the sound propagation which we think is worth reporting.

In section \ref{sec_tunneling} we discuss the ``finite amplitude" instability  
developing via more subtle ``tunneling" scenario. In this case both phases retain the shape of local minima in
the functional space, however separated by a relatively small or ``penetrable" barrier.

      
     \subsection{The holographic models}

     Pure gauge $SU(N_c)$ theory is known to have the first order deconfinement phase transition for $N_c>2$. While $u,d,s$ quarks with physical masses seem to
  change it to a crossover, the QCD thermodynamics still is quite close to the first-order behavior.   Rather detailed information about
thermodynamical quantities comes from  the first principles via the lattice gauge theory. Unfortunately, this approach does require 
analytic continuation into the Euclidean time, and thus it has very limited capabilities for predicting
real-time dynamics such as  dissipative phenomena. One also cannot look for instabilities and their dynamics in the Euclidean setting.

   Another tool we have to get to the strongly coupled regime are 
  the  holographic models. A limitation of those is that they are used mostly  in the large $N_c$ limit.
However they can be used both in Euclidean and real time settings. In fact their crucial success  has been the
calculation of the near-equilibrium bulk and shear viscosities. 
The famous prediction for the latter  $\eta/s=1/4\pi$   is within the factor of 2 from the phenomenological value.

We will not have place here to give introduction to those models or their historical development:
the interested reader can consult e.g.  Ref \cite{GKN}.  
We directly introduce the notations and the action, for the generic dilaton-gravity model is
\begin{eqnarray}\label{Lag}
S=-\frac{1}{16\pi G_5}\int d^5x\sqrt{G}\(R-\frac{4}{3}\(\del\phi\)^2-{\cal V}(\phi)\) \nonumber \\
+\frac{1}{8\pi G_5}\int_\del d^4x\sqrt{H}K.
\end{eqnarray}
where the last term is the Gibbons-Hawking term of the boundary, with $H,K$ being the corresponding induced metric and extrinsic curvature. 
Depending the choice of dilaton potential, different groups have succeeded in matching to different aspects of real world QCD \cite{GKLN,MPV}. In this work, we will use the specific model in \cite{MPV}.

The corresponding equations of motion admits two solutions, which we will call the low-T (``hadronic") and the high-T (``blackhole" or ``QGP"), respectively.
The low-T solution is described by the metric and scalar field $z$ depended on the holographic coordinate $z$ 
\begin{align}\label{vac}
ds^2&=b_0(z)^2(-dt^2+d{\vec x}^2+dz^2),\, \af=e^{\phi}=\af_0(z)
\end{align}
while the high-T solution has the blackhole form
\begin{align}\label{bh}
ds^2&=b(z)^2(-f(z)dt^2+d{\vec x}^2+{dz^2 / f(z)}),\, \af=\af(z)
\end{align}
The low-T solution exists for any temporal extension given by the inverse temperature. The high-T solution exists only above a minimum temperature $T_{min}$. Furthermore, it becomes thermodynamically favorable above a critical temperature $T_c$, as seen from the calculation of its free energy.  So, below $T_c$ the blackhole solution is $metastable$ and can in principle be realized by super-cooling.

Equilibrium coexistence between the QGP cluster (called in this paper the plasma-ball) and hadronic matter has been first discussed by 
Aharony, Minwalla and Wiseman \cite{Aharony:2005bm}, who had formulated  the problem in holographic setting
and had solved it in the  domain wall case, of a flat boundary between the two phases.
This solution is naturally  stable only at $T=T_c$, as there is no force acting on the wall.
 Its energy per area -- the surface tension -- has been evaluated in \cite{Aharony:2005bm}.

\section{Overcooled high-T phase and Quasi-Normal Modes in the sound channel}
\label{sec_modes}
 
The question is whether at certain degree of supercooling, the free energy minimum may turn into a maximum,
 at least in some direction in functional space, so that
 an instability may develop. We will study those triggered in a channel involving the dilaton. This is the sound channel, which mixes components of stress tensor with the gluon condensate in the holographic QCD model. The perturbation of bulk fields are parametrized as:
\begin{eqnarray}
ds^2&=&b(z)^2(-f(z)dt^2+d{\vec x}^2+dz^2/f(z)+h_{\mu\nu}dx^\mu dx^\nu), \nonumber \\
\af&\to&\af+\dlt\af .
\end{eqnarray}
In the axial gauge for the metric perturbation, we have $\mu=t,x_1,x_2,x_3$. Taking $h_{\mu\nu}$ and $\dlt\af$ to be plane wave forms, with spatial momentum lie along $x_3$ direction. The sound channel involves the following metric components
\begin{eqnarray}
h_{00}, h_{33}, h_{03}, h_{aa}=(h_{11}+h_{22})/2,
\end{eqnarray}
and the perturbation of the dilaton $\dlt\af$. The EOM have been previously derived in \cite{springer}. We sketch the derivation here for completeness. It is convenient to study the perturbation in terms of the following gauge invariant combinations:
\begin{align}\label{combo}
Z_0&=k^2h_{00}+2k\omg h_{03}+\omg h_{33}-\(\omg^2-k^2\(f+\frac{b f'}{2b'}\)\)h_{aa}, \nonumber \\
Z_\af&=\frac{\dlt\af}{\af}-\frac{\af'b}{2\af b'}h_{aa}.
\end{align}
The EOMs can be derived from the linearized Einstein equation and dilaton equation,
\begin{eqnarray}\label{eom_Z}
Z_0''+P(\af(z),b(z),f(z))Z_0'+Q(\af(z),b(z),f(z))Z_0  \nonumber \\
+R(\af(z),b(z),f(z))Z_\af=0 \nonumber \\
Z_\af''+K(\af(z),b(z),f(z))Z_\af'+L(\af(z),b(z),f(z))Z_\af  \nonumber \\
+M(\af(z),b(z),f(z))Z_0'+N(\af(z),b(z),f(z))Z_0=0. 
\end{eqnarray}
The coefficients are functionals of the background solution. The explicit forms are lengthy and will not be shown here. The fluctuations of $Z_0$ and $Z_\af$ can be studied by integrating \eqref{eom_Z} from the horizon. Near the horizon, the gauge invariant combinations satisfying infalling boundary condition have the expansion
\begin{eqnarray}
Z_0&=(z_h-z)^{-\frac{i\omg}{4\pi T}}\(c_0+c_1(z_h-z)+\cdots\), \no
Z_\af&=(z_h-z)^{-\frac{i\omg}{4\pi T}}\(d_0+d_1(z_h-z)+\cdots\),
\end{eqnarray}
where analytic form of $c_1$, $d_1$ etc can be expressed as functions of $c_0$ and $d_0$ only. The coefficients $c_0$ and $d_0$ are arbitrary. QNM of the coupled system corresponds to the fluctuations that are infalling at the horizon and vanishing at the boundary. 
The latter condition means
\begin{eqnarray}
\lim_{z\to0}Z_0 = 0, \lim_{z\to0}Z_\af\frac{\af b'}{\af' b} = 0.
\end{eqnarray}
The QNM are realized only for a particular ratio of $c_0$ and $d_0$. The condition can be formulated as
\begin{eqnarray}
\lim_{z\to0}Z_0^{(1)}Z_\af^{(2)}\frac{\af b'}{\af' b}-Z_0^{(2)}Z_\af^{(1)}\frac{\af b'}{\af' b} = 0,
\end{eqnarray}
where the supscript indicates a pair of linearly independent solutions. In practice, we choose $c_0=0,d_0=1$ and $c_0=1,d_0=0$ for the pair. 

\begin{figure}
\includegraphics[width=0.4\textwidth]{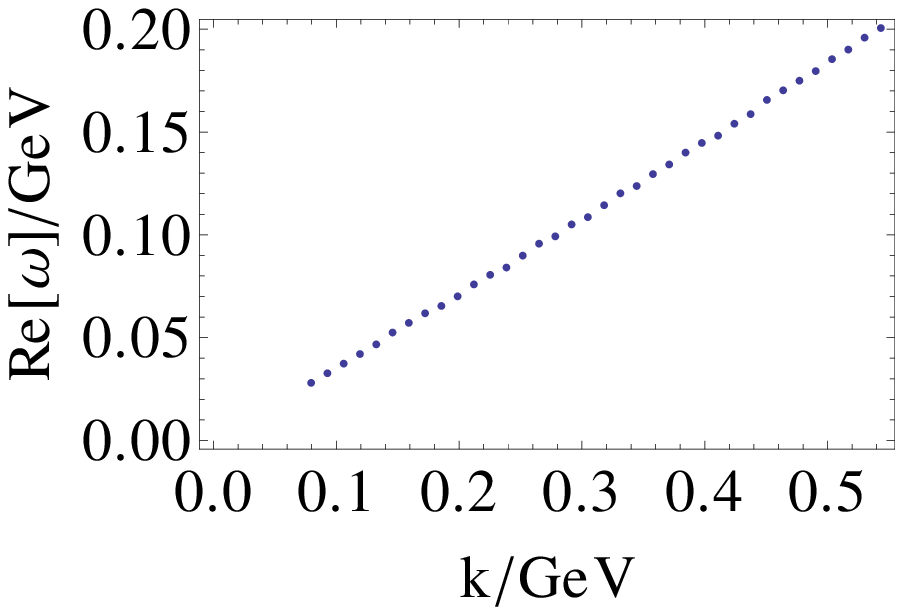}
\includegraphics[width=0.4\textwidth]{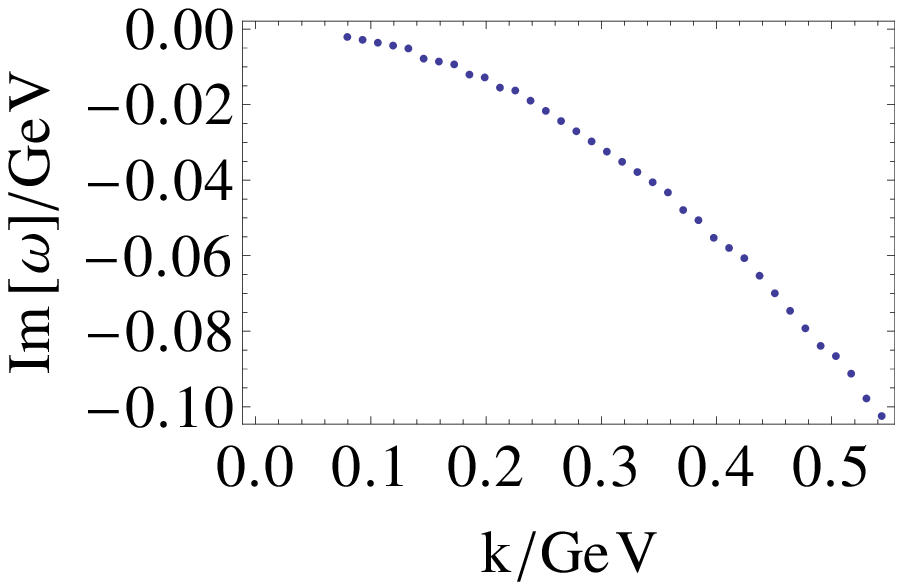}
\caption{\label{qnm12}QNM for super-cooled blackhole with $\af_h=1.2$.}
\end{figure}

\begin{figure}
\includegraphics[width=0.4\textwidth]{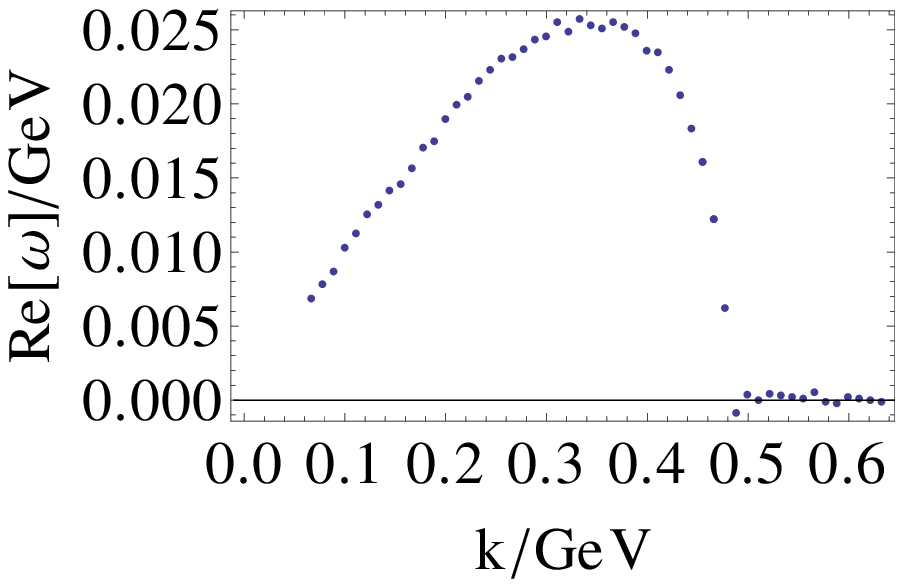}
\includegraphics[width=0.4\textwidth]{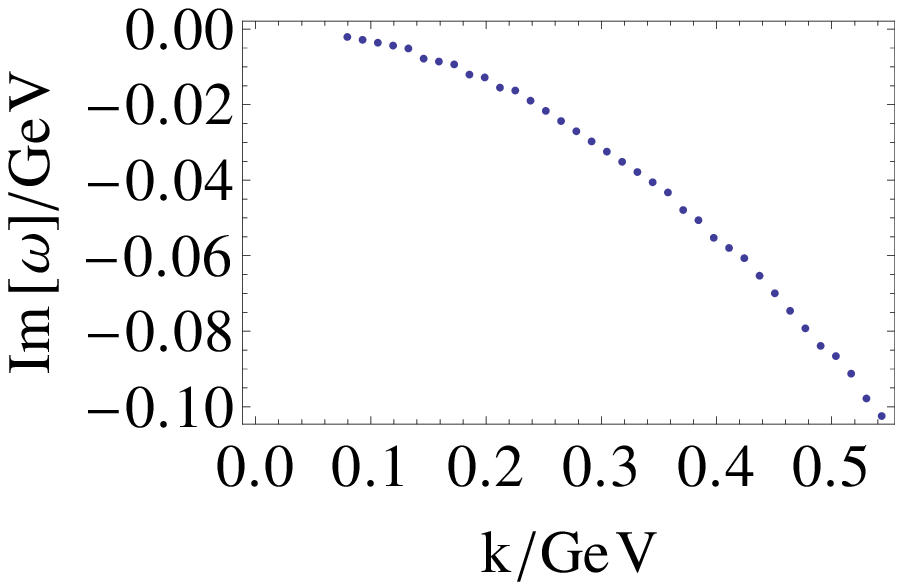}
\caption{\label{qnm16}QNM for super-cooled blackhole with $\af_h=1.6$. The real part of QNM are consistent with zero beyond the critical momentum.}
\end{figure}

Before presenting our numerical results for the QNM search, we explain a useful trick in the numerical procedure. We first recall that the numerical solution of the blackhole background is obtained first by starting with arbitrary integration constants at the horizon and then applying a set of rescalings to obtain the true numerical solution. Explicitly, the equations for the background 
\begin{eqnarray}\label{eom_bg}
\left\{\begin{array}{l}
W'=\frac{16}{9}bW^2-\frac{1}{f}\(Wf'-\frac{3}{4}b{\cal V}\),\\
b'=-\frac{4}{9}b^2W,\\
\af'=\af\sqrt{bW'}\\
f''=\frac{4}{3}f'bW,
\end{array}
\right.
\end{eqnarray}
are invariant under the following rescalings
\begin{eqnarray}\label{rescale}
\begin{array}{l}
W(z)\to  W(z+\xi),\;\; b(z)\to b(z+\xi),\;\; \af(z)\to \af(z+\xi),\quad \nonumber \\  f(z)\to f(z+\xi). \\
W(z)\to  W(z)\sqrt{\dlt_f},\;\; b(z)\to b(z)/\sqrt{\dlt_f},\;\; \af(z)\to \af(z),\nonumber \\ f(z)\to f(z)/\dlt_f. \\
W(z)\to  W(z\dlt_b),\;\; b(z)\to b(z\dlt_b)\dlt_b,\;\; \af\to\af(z\dlt_b), \nonumber \\ f(z)\to f(z\dlt_b).
\end{array} 
\end{eqnarray}
These are  simply the manifestation of isometries in the particular background. The first line of \eqref{rescale} is just trivial shift in $z$ coordinate. The second is inhomogeneous rescaling in $t$ and $x$. The third line is homogeneous rescaling in $t$, $x$ and $z$. They have a straight-forward generalization in the presence of metric and dilaton perturbations, which we list as follows
\begin{eqnarray}\label{scale_tx}
\begin{array}{l}
t\to t\dlt_f,\;\; x\to x\sqrt{\dlt_f},\;\; h_{00}\to h_{00}/\sqrt{\dlt_f}, h_{33}\to h_{33},
\\ h_{03}\to h_{03}/\sqrt{\dlt_f}, h_{aa}\to h_{aa}, \dlt\af\to \dlt\af. \\ \nonumber
t\to  t/\sqrt{\dlt_b},\;\; x\to x/\sqrt{\dlt_b},\;\; z\to z/\sqrt{\dlt_b}, h_{00}\to h_{00},\;\; 
 \\ h_{33}\to h_{33}, 
  h_{03}\to h_{03}, h_{aa}\to h_{aa},\;\; \dlt\af\to \dlt\af.  \nonumber
\end{array} \end{eqnarray}
They correspond to the following scaling rule for $\omg$ and $k$:
\begin{eqnarray}\label{scale_omgk}
&\omg\to\omg/\dlt_f,\;\; k\to k/\sqrt{\dlt_f}. \\ \nonumber
&\omg\to\omg\dlt_b,\;\; k\to k\dlt_b.
\end{eqnarray}
With these in mind, we can work with blackhole background obtained with arbitrary integration constants and solve the fluctuation equations for the corresponding QNM. The true QNM can be obtained in the end by applying the rescaling rule \eqref{scale_omgk}.
We show the resulting QNM in the super-cooled blackhole phase corresponding to $\af_h=1.2$ and $\af_h=1.6$ in 
Figures \ref{qnm12} and \ref{qnm16}, with the latter corresponding to blackhole with the lowest temperature $T_{min}$. 

We see no sign of QNM developing a positive imaginary part: thus we conclude that {\em there is no instability of the sound mode}.
However, we do find quite a remarkable phenomenon in form of radical changes in the sound propagation. While the former plots with $\af_h=1.2$
show the conventional sound -- with real part approximately linearly rising with $k$ and imaginary part approximately quadratic in $k$ 
--  the solution with $\af_h=1.6$ is entirely different. Above certain momentum 
the sound {\em stops propagating completely},  $Re(\omega)\rightarrow 0$, and the mode becomes purely dissipative. On the other hand the slope $\del Im(\omg)/\del k$ has a discontinuity at this critical momentum.
Such regime is of interest, and as far as we know, it is observed for the first time. A similar behavior in the diffusion mode for normal QGP at finite momentum has been found in \cite{landsteiner}, where the purely dissipative diffusion mode starts to propagate with $Re(\omg)>0$ beyond a certain critical momentum. In that case, the behavior originates in the crossing of the diffusion mode with the lowest non-hydrodynamic mode. It would be interesting to see if the stoppage of sound mode in supercooled black hole has similar origin.


\section{Collapse of the QGP clusters}
\label{sec_collapse}
  We now jump to the end of the inhomogeneous phase, and consider the fate of the high-$T$ phase, which in our case is generically called QGP. 
  
  For very large clusters one can return to the flat domain wall solution of  Aharony et al \cite{Aharony:2005bm},
  take $T<Tc$ and calculate the force experienced by the domain wall (which naturally acts in the direction of the QGP side).
  Also, since the surface tension is known from their work, one can immediately estimate the size of the so called critical bubble,
  for which the gain in the volume and loss in surface free energy compensate each other
  \be R_*= {\sigma \over |\delta F| } \ee  
  Smaller clusters, with $R<R_*$ would disappear by classical shrinkage, while the large ones are metastable and require tunneling. 
  In macroscopic physics, in which one can tune conditions to be very close to critical and $|\delta F| $ very small, such critical
  bubbles are macroscopically large and the  treatment outlined above is obviously justified.

 Unfortunately, in heavy ion collisions we do not have very large systems. Therefore the explosions are rather rapid and
    time between $T_c$ and freezeout is quite limited.  Since this temperature is not very close to $T_c$ and thus  there is no parameter available which makes the critical cluster large.
    More specifically,
     only the final hadrons emerging at the ``freeze out" temperature in the range $T_f=120-150\, MeV$  are actually observed,
  substantially lower than the critical one $T_c\approx 170 \, MeV$. 
  
  One may wary that since $T_f$ is not that close to $T_c$, all critical phenomena will be erased. But -- as we already noted in the Introduction --
   two-particle rapidity correlation functions does show existence of certain clusters, from which
  observed hadrons come. The multiplicity per cluster is in the range 5-10, which is not a large number.
 If clusters are not large, separation of their free energy into the volume, surface, curvature and so on terms makes little sense.
 Therefore, one needs to address the issue of ``mesoscopic" clusters, of the size comparable to the basic correlation scales.
 
   One possible --albeit technically challenging -- way to proceed would be to generalize the solution of  Aharony et al \cite{Aharony:2005bm}
    from flat to spherical geometry, looking for classically stable solutions with the critical size, as a function of $T$. 
Unfortunately, even at $T_c$ and for flat domain wall considered in that paper, the solution is complicated by the fact
that in holography the domain wall solution  depends not only on the coordinate normal to the wall, $x$, but also on the holographic
coordinate we will now generically call $z$. In fact    Aharony et al  were only able to get the solution 
in coordinates in which they had shown certain symmetry between the arguments, on basis of which one
can get to a kind of radial coordinate of the kind of $\sqrt{x^2+z^2}$ and make the problem one-dimensional. 
  For a spherical cluster the spatial radial coordinate $r$ does not enter the metric like the flat $x$ from the beginning, and no
   symmetry with the holographic coordinate $z$ of the kind can be expected, and thus one has to face equations in partial derivatives with different dependence on $r,z$ variables.  
    
   In this paper we  propose a way around this difficulty, which is not strict but allows certain practical progress. The idea
 is to consider field configurations depending on only $one$ variable, the holographic one,  incorporating {\em two different phases}
 in the UV and IR directions.
We will call it a domain-wall (DW) background. It interpolates a high-T
blackhole (BH) solution in the UV and a low-T thermal gas (TG) solution in the IR. If the width of the wall is small, the background can be approximated by a blackhole patch and a thermal gas patch joined by a interface. This corresponds to the thin-wall limit \cite{deluccia}. While in general the metric and dilaton can have finite jumps across the interface, we will restrict ourselves to the case where they are continuous. The approximate background satisfies Einstein equation everywhere except on the interface, where it deviate from the true solution. We will refer to the interface as {\em membrane} in the following. (Thus the setting resembles the one in our paper 
\cite{Lin:2008rw} in which the dynamics of the interface should be determined from Israel junction condition. Similar setting has also been considered in global AdS space for studying decay of unstable CFT state \cite{BR}). We stress however, that -- unlike  \cite{Lin:2008rw,BR} -- we don't introduce any external source to the gravity-dilaton fields.   
Our background is homogeneous and isotropic in spatial coordinates of field theory, the position of the membrane can only be a function of time. We will parametrize the position of the membrane by $z=\Lam$($z_0=\Lam_0$) and derive an action for its evolution. The dynamics of the membrane will be determined from the action. Decay from supercooled BH to TG will happen if the membrane moves toward the boundary, enlarging the TG patch. 
By substituting our approximate background into the action of the model, we get effective action of motion for the membrane,
which would describe the cluster collapse at appropriate parameters. 


The action per unit 4-volume is obtained from the renormalized action of the background \footnote{See \cite{GKLN} for details of this renormalization procedure}.
\begin{eqnarray}\label{S_full}
\frac{S}{\bt\text{Vol}}=\lim_{\eps,\tilde{\eps}\to0}\(S^{DW}(\eps)-S^{TG}(\tilde{\eps})\),
\end{eqnarray}
The action of the DW contains three contributions: the BH patch $0<z<\Lam^-$, the TG patch $z_0>\Lam_0^+$ and the membrane $z>\Lam^-$, $z_0<\Lam_0^+$.
The contributions to the action from the bulk BH and TG patches, which we denote as $S_1^{DW}$ are easily obtained as
\begin{align}\label{S_dw}
\frac{{S}_1^{DW}(\eps)}{\bt\text{Vol}} 
&=-\frac{1}{16\pi G_5}\int_\eps^\Lam dz\sqrt{G}\(R-\frac{4}{3}(\del\phi)^2-{\cal V}(\phi)\) \no
&+\frac{1}{8\pi G_5}\sqrt{H}K|_{z=\eps} \no 
&-\frac{1}{16\pi G_5}\int_{\Lam_0}^\infty dz\sqrt{G}\(R-\frac{4}{3}(\del\phi)^2-{\cal V}(\phi)\),
\end{align}
where the TG action that appears in \eqref{S_full} reads
\begin{align}
\frac{S^{TG}(\tilde\eps)}{\bt\text{Vol}} 
&=-\frac{1}{16\pi G_5}\int_{\tilde\eps}^\infty dz\sqrt{G}\(R-\frac{4}{3}(\del\phi)^2-{\cal V}(\phi)\) \no
&+\frac{1}{8\pi G_5}\sqrt{H}K|_{z=\tilde\eps}\, . 
\end{align}
Note that we don't include Gibbon-Hawking terms at $z=\Lam$ and $z_0=\Lam_0$ in (\ref{S_dw}). The reason is that they will be cancelled by corresponding terms in the membrane action obtained by integrating from $z=\Lam^-$ to $z_0=\Lam_0^+$.
Using \eqref{S_dw} and following \cite{GKLN}, this particular contribution to the difference in \eqref{S_full}, $S^{DW}_1-S^{TG}$ evaluates to
\begin{align}\label{S_bulk}
&\frac{S_1}{\bt\text{Vol}}= \frac{1}{16\pi G_5}\bigg[\no
&\lim_{\eps\to0}\(6b^2(\eps)b'(\eps)f(\eps)+b(\eps)^3f'(\eps)-6b_0'(\tilde{\eps})\frac{b(\eps)^4\sqrt{f(\eps)}}{b_0^2(\tilde{\eps})}\) \no
&+2f(\Lam)b(\Lam)^2b'(\Lam)-2b_0(\Lam_0)^2b_0'(\Lam_0)\frac{\bt_0Vol_0}{\bt Vol}\bigg],
\end{align}
where the first line is identified as the BH free energy density. The very last term is due to the TG patch, while the rest is contribution from the BH patch. The dependence on $\Lam$ is contained in the second line. 

Here $\Lam$ and $\Lam_0$ are coordinates of the membrane from the BH and TG solutions respectively. They are related by the continuity condition of the dilaton
\begin{eqnarray}\label{dilaton_cont}
\af(z=\Lam)=\af_0(z_0=\Lam_0).
\end{eqnarray}
 The ratio of the 3-volume in the last term is easily obtained by matching the spatial metric components on the membrane, with all quantities evaluated on the membrane:
\begin{align}
b^2dx^2=b_0^2dx_0^2\; \Rightarrow\;\frac{Vol_0}{Vol}=\frac{d^3x_0}{d^3x}=\frac{b^3}{b_0^3}.
\end{align}
The ratio of the temporal extension is obtained by matching the metric components along the trajectory of the membrane. The result will necessarily depend on the moving velocity of the membrane.
\begin{align}
b^2(-fdt^2+dz^2/f)=b_0^2(-dt_0^2+dz_0^2).
\end{align}
Using $dz=\dot\Lam dt$, $dz_0=\dot\Lam_0dt_0$ on the membrane trajectory where dot denotes the time derivative and $\af'd\Lam=\af_0'd\Lam_0$ from the continuity of the dilaton, we obtain
\begin{align}
\frac{\beta_0}{\beta}=\frac{dt_0}{dt}=\frac{\sqrt{b^2f-\(\frac{b^2}{f}-\frac{\af'{}^2}{\af_0'{}^2}b_0^2\)\dot\Lam^2}}{b_0}.
\end{align} 
In the limit $\dot\Lam\to0$, $\frac{\bt_0}{\bt}\to\frac{b\sqrt{f}}{b_0}$, recovering the stationary boundary result as in \cite{GKLN}. 
 Collecting everything, \eqref{S_bulk} takes the following form:
\begin{align}\label{s1}
&\frac{S_1}{\bt Vol}=F+ \no
&\frac{1}{16\pi G_5}\bigg[2fb^2b'-\frac{2b_0'b^3\sqrt{b^2f-\(\frac{b^2}{f}-\frac{\af'{}^2}{\af_0'{}^2}b_0^2\)\dot\Lam^2}}{b_0^2}\bigg],
\end{align}
where the quantities with(without) subscript $0$ are evaluated at $\Lam$($\Lam_0$). The first term in (\ref{s1}) is the renormalized free energy of the blackhole background \cite{GKLN}. Without dependence on $\dot\Lam$, \eqref{s1} may be viewed as a potential energy.

The less obvious contribution from the membrane is obtained as follows: Recall that the dilaton and metric are continuous across the membrane. Their first derivatives however can have finite jumps and second derivatives can have delta functions. The Ricci scalar in \eqref{Lag} contains second derivatives in metric components, which upon integration across the membrane from $z=\Lam^-$ to $z_0=\Lam_0^+$, will give finite contribution to the action. As will be clear soon, it gives contribution to both kinetic energy and potential energy. To apply this method, we need to require the metric components are continuous. This is a more stringent condition than the continuity of metric on the membrane. To this end, we rescale the coordinates $t,x,z$ in the TG solution so that the metric components are continuous across the membrane:
\begin{align}
ds^2&=b_0(z_0)^2(-dt_0^2+dx_0^2+dz_0^2)\rightarrow \no
ds^2&=b_0(z_0)^2g(\Lam,\Lam_0)(-f(\Lam)dt^2+dx^2+d{\bar z}_0^2/f(\Lam)),
\end{align}
where $g(\Lam,\Lam_0)\equiv \frac{b(\Lam)^2}{b_0(\Lam_0)^2}$.
The rescaled radial coordinate $\bar{z}_0$ and the original one $z_0$ are related by
\begin{eqnarray}
b_0(\Lam_0)dz_0=\frac{b(\Lam)}{\sqrt{f(\Lam)}}d\bar{z}_0,
\end{eqnarray}
from which it follows the relation on the derivatives
\begin{eqnarray}
\del_{\bar{z}_0}=\frac{b(\Lam)}{b_0(\Lam_0)\sqrt{f(\Lam)}}\del_{z_0}.
\end{eqnarray}
Now we can write the metric components for BH and TG solutions in a unified way
\begin{align}\label{metric_unify}
ds^2&=g_{tt}(t,\tz)dt^2+g_{xx}(t,\tz)dx^2+g_{zz}(t,\tz)d\tz^2 \no
\text{with}& \no
g_{tt}&=-b^2(\tz)f(\tz)\theta(\Lam-\tz)-b_0(z_0)^2g(\Lam,\Lam_0)f(\Lam)\theta(\tz-\Lam) \no
g_{xx}&=b^2(\tz)\theta(\Lam-\tz)+b_0(z_0)^2g(\Lam,\Lam_0)\theta(\tz-\Lam) \no
g_{zz}&=b^2(\tz)/f(\tz)\theta(\Lam-\tz)+b_0(z_0)^2g(\Lam,\Lam_0)/f(\Lam)\theta(\tz-\Lam).
\end{align}
We have defined
\begin{eqnarray}
\tz &=\left\{\begin{array}{ll}
z, & \tz<\Lam\\
\bar{z}_0, & \tz>\Lam
\end{array}
\right ..
\end{eqnarray}
We stress again that $\Lam$ and $\Lam_0$ are related by \eqref{dilaton_cont}, and both are implicitly $t$-dependent.
Substituting contributions from \eqref{metric_unify} in the Ricci scalar, we find the following second derivatives contribute
\begin{align}\label{metric_dd}
\frac{\del^2g_{tt}}{\del\tz^2}&=\(2bb'f+b^2f'-2b_0'\frac{b^3}{b_0^2}\sqrt{f}\)\dlt(\tz-\Lam) \no
\frac{\del^2g_{xx}}{\del\tz^2}&=\(-2bb'+2b_0'\frac{b^3}{b_0^2\sqrt{f}}\)\dlt(\tz-\Lam) \no
\frac{\del^2g_{xx}}{\del t^2}&=-\(2bb'-\frac{2b^2b_0'\af'}{b_0\af_0'}\)\dot{\Lam}^2\dlt(\tz-\Lam) \no
\frac{\del^2g_{zz}}{\del t^2}&=-\(\frac{2bb'}{f}-\frac{2b^2b_0'\af'}{b_0f\af_0'}-\frac{b^2f'}{f^2}\)\dot{\Lam}^2\dlt(\tz-\Lam).
\end{align}
In the above, we have only kept terms proportional to the delta function. The other terms vanish upon infinitesimal integration across the membrane. All quantities with(without) indices $0$ are evaluated at $\Lam$($\Lam_0$). Dot means derivative with respect to $t$. Combining \eqref{metric_dd} all in all we obtain the following contribution from the membrane as
\begin{align}\label{s2}
\frac{S_2}{\bt Vol}&=\frac{1}{16\pi G_5}\bigg[\(-8b^2fb'-b^3f'+\frac{8b^4b_0'\sqrt{f}}{b_0^2}\) \no
&+\(\frac{8b^2b'}{f}-\frac{b^3f'}{f^2}-\frac{8b^3b_0'\af'}{b_0f\af_0'}\)\dot\Lam^2\bigg].
\end{align}
Obviously terms containing $\dot{\Lam}^2$ correspond to the kinetic energy while the rest gives additional contribution to the potential energy. Note one important fact: all the terms with the second time derivative of $\Lambda$ automatically cancel out, basically because of continuity of the metric components.

Summing over \eqref{s1} and \eqref{s2}, we obtain the following expression for the action for DW background:
\begin{align}\label{s12}
\frac{1}{\bt\text{Vol}}{S}&=F+\frac{1}{16\pi G_5}\bigg[\(-6b^2fb'-b^3f'+\frac{8b^4b_0'\sqrt{f}}{b_0^2}\) \no
&+\(\frac{8b^2b'}{f}-\frac{b^3f'}{f^2}-\frac{8b^3b_0'\af'}{b_0f\af_0'}\)\dot\Lam^2 \no
&-\frac{2b_0'b^3\sqrt{b^2f-\(\frac{b^2}{f}-\frac{\af'{}^2}{\af_0'{}^2}b_0^2\)\dot\Lam^2}}{b_0^2}\bigg].
\end{align}
Again all quantities with(without) indices $0$ are evaluated at  $\Lam$($\Lam_0$). The first line is potential energy and the second line is kinetic energy. The third line is noncanonical. If we formally expand in $\dot\Lam^2$, the zeroth order term gives additional contribution to potential energy and first order term gives correction to the kinetic energy. It is useful to note the existence of conserved energy due to the absence of explicit $t$-dependence of the action \eqref{s12}. Legendre transform of \eqref{s12} gives the following conserved energy:
\begin{align}\label{h12}
\frac{H}{\bt Vol}&=-F-\frac{1}{16\pi G_5}\bigg[\(-6b^2fb'-b^3f'+\frac{8b^4b_0'\sqrt{f}}{b_0^2}\) \no
&-\(\frac{8b^2b'}{f}-\frac{b^3f'}{f^2}-\frac{8b^3b_0'\af'}{b_0f\af_0'}\)\dot\Lam^2 \no
&-\frac{2b_0'b^5f}{b_0^2}\frac{1}{\sqrt{b^2f-\(\frac{b^2}{f}-\frac{\af'{}^2}{\af_0'{}^2}b_0^2\)\dot\Lam^2}}\bigg].
\end{align}

%

\subsection{Quantum tunneling of membrane in DW background}\label{sec_tunneling}
\begin{figure}
\includegraphics[width=0.4\textwidth]{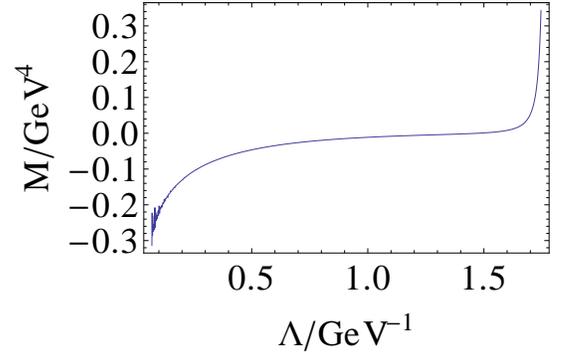}
\caption{\label{fig_M} $M$ as a function of $\Lam$ for supercooled blackholes with $\af_h=1.2$. The left edges near $\Lam=0$ are noisy as it involves a subtraction of infinity from infinity. One analytically finds that $M$ logarithmically diverges as $- \log\Lam$ as $\Lam\to 0$. The right boundary is $\Lam=z_h$.}
\end{figure}

\begin{figure}
\includegraphics[width=0.4\textwidth]{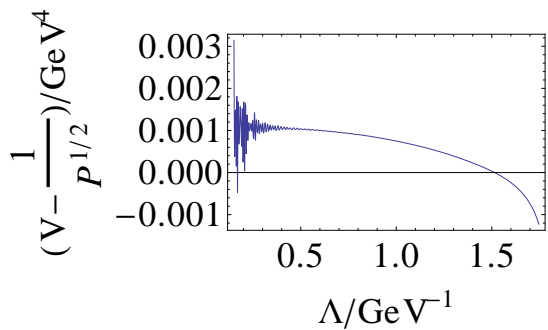}
\caption{\label{fig_V}. $V-\frac{1}{\sqrt{P(\Lam)}}$ as a function of $\Lam$ for the supercooled blackhole with $\af_h=1.2$. The left edges near $\Lam=0$ are noisy as it involves a subtraction of infinity from infinity. One analytically finds the limit value as $F$ as $\Lam\to 0$. The right boundary is $\Lam=z_h$.}
\end{figure}

Now we are ready to study the dynamics of the membrane.
The functions appearing in \eqref{s12} need to be determined numerically from the background BH and TG solutions. Here we present the numerical results of those functions and discuss some of the analytical features. We first define 
\begin{align}\label{sh_pk}
\frac{S}{\bt\text{Vol}}&=F-V(\Lam)+M(\Lam)\dot\Lam^2+\frac{\sqrt{P(\Lam)-Q(\Lam)\dot\Lam^2}}{P(\Lam)} \no
\frac{H}{\bt\text{Vol}}&=-F+V(\Lam)+M(\Lam)\dot\Lam^2-\frac{1}{\sqrt{P(\Lam)-Q(\Lam)\dot\Lam^2}},
\end{align}
where
\begin{align}\label{vmpq}
16\pi G_5V(\Lam)&=-\(-6b^2fb'-b^3f'+\frac{8b^4b_0'\sqrt{f}}{b_0^2}\), \no
16\pi G_5M(\Lam)&=\(\frac{8b^2b'}{f}-\frac{b^3f'}{f^2}-\frac{8b^3b_0'\af'}{b_0f\af_0'}\), \no
\(16\pi G_5\)^{-2}P(\Lam)&=\frac{b_0^4}{4b^8fb_0'^2}, \no
\(16\pi G_5\)^{-2}Q(\Lam)&=\frac{b_0^4\(\frac{b^2}{f}-\frac{\af'{}^2}{\af_0'{}^2}b_0^2\)}{4b^{10}f^2b_0'^2}.
\end{align}
In Figures \ref{fig_M} and \ref{fig_V}, we present numerical results on mass function $M(\Lam)$ and $V(\Lam)-\frac{1}{\sqrt{P(\Lam)}}$. The 5D gravitational constant appearing in the action is given by $\frac{1}{16\pi G_5}=\frac{N_c^2-1}{45\pi^2 l^3}$, which evaluates to $2\times 10^{-4}GeV^3$ with $l=4.389GeV^{-1}$ fixed in \cite{MPV}.  We note a peculiar behavior of the mass function: it is positive near the horizon and negative near the boundary, with a sign change in the middle. We denote the zero of $M(\Lam)$ by $\Lam_m$. The physical meaning of $V(\Lam)-\frac{1}{\sqrt{P(\Lam)}}$ is the Hamiltonian density $\frac{H}{\bt\text{Vol}}$ evaluated at $\dot\Lam=0$. The numerical results on $V(\Lam)-\frac{1}{\sqrt{P(\Lam)}}$ is noisy near $\Lam=0$, as it follows from numerically subtracting an infinity from another. However, we have analytical knowledge of the limit $\Lam\to0$:
\begin{align}
&V(\Lam)-\frac{1}{\sqrt{P(\Lam)}}=\frac{1}{16\pi G_5}\(6b^2fb'+b^3f'-\frac{6b^4b_0'\sqrt{f}}{b_0^2}\) \no
&\to { F}.
\end{align}
Therefore, $H(\Lam\to0,\dot\Lam=0)$ is positive for $T<T_c$ and negative for $T>T_c$. The opposite limit $\Lam\to z_h$ gives a negative value because $f(z_h)=0$ and $f'(z_h)<0$. Using the generic expressions in \cite{GKLN} one finds
\be\label{hir} 
V(\Lam)-\frac{1}{\sqrt{P(\Lam)}} \to -ST, \qquad  \Lam\to z_h\, ,
\ee
 where $S$ is the entropy density of the black-hole and T is temperature. 
As will be clear soon, this feature is in favor of tunneling in the supercooled black hole, but suppress tunneling in the normal black hole. Furthermore, the positivity of $P(\Lam)$ is obvious from its expression. One also observes numerically that $Q(\Lam)$ is positive up to region very close to the boundary, where numerical noise becomes significant. In fact one can prove that $Q>0$ for the black-hole in the entire range of $\Lam$ using the generic equations for the dilatonic black-holes presented in \cite{GKLN}.   

We proceed by solving for $\dot\Lam$ from the conserved energy. Since we are considering a process starting from a stationary supercooled BH, the conserved energy follows from Legendre transform of supercooled blackhole action. One has to note however that in the limit $\Lam\to z_h$ the contribution of the membrane action $S_2$ (\ref{s2}) should be subtracted. The rest, that is, the contribution $S_1$ (\ref{s1}) vanishes provided that $\dot\Lam^2$ vanishes faster than linear near the horizon. This is indeed satisfied in the solutions we find below. Motivated by this, we set the Hamiltonian to the value one obtains by Legendre transforming $S_1$ and evaluating at $z_h$. This gives $H=-F$. The equation of motion for the membrane is then:    
  \begin{eqnarray}\label{eq_dotlam}
\!\!\!\!\!\!\!\!\!\!\!\!A(\Lam,\dot\Lam)\!\equiv \!V(\Lam)+\!M(\Lam)\dot\Lam^2\!-\!\frac{1}{\sqrt{P(\Lam)-Q(\Lam)\dot\Lam^2}}=0.
\end{eqnarray}
For a given $\Lam$, the solution to $\dot\Lam^2$ can be positive, negative or complex. A positive solution means that the classical path is possible and the membrane can move spontaneously. In our scenario it moves toward the boundary, squeezing the QGP phase. A negative solution corresponds to a  purely imaginary $\dot\Lam$. The realization of such  a solution requires quantum tunneling. The classic tunneling description of metastable bubbles has been proposed via the so called ``bounce" classical solution long ago, by 
Kobzarev et al \cite{Kobzarev:1974cp} and Coleman \cite{Coleman:1977py,deluccia}.
The basic idea is to Wick-rotate time to imaginary axis, thus the phase shift $e^{iS}$ becomes $e^{-S}$, giving the tunneling probability. Note that for negative $\dot\Lam^2$, $S$ is always real. This can be generalized to include complex $\dot\Lam^2$ as well, in which case $S$ can also be complex. The phase shift can be expressed as
\begin{align}
i\int dtd^3x\frac{S}{\bt\text{Vol}}=i\int \frac{d\Lam d^3x}{\dot\Lam}\frac{S}{\bt\text{Vol}}.
\end{align}
The imaginary part of the integrated action $\int \frac{d\Lam d^3x}{\dot\Lam}\frac{S}{\bt\text{Vol}}$ is related to tunneling probability.

It is instructive to discuss the tunneling window using general properties of functions $V(\Lam)$, $M(\Lam)$, $P(\Lam)$ and $Q(\Lam)$. 
We have shown that as $\Lam\to0$, $A(\Lam,\dot\Lam=0)\to F$, thus is positive for $T<T_c$, and it tends to a negative value $-ST$ as $\Lam\to z_h$. Furthermore, we find numerically it is a monotonous function. Therefore $A(\Lam,\dot\Lam=0)$ has a single zero, which we denote as $\Lam_h$. For sufficient low T, $\Lam_m<\Lam_h$. When $0<\Lam<\Lam_m$, we have $M(\Lam)<0$ and because $P(\Lam)>0$, $Q(\Lam)>0$, $A(\Lam,\dot\Lam^2)$ is a monotonously decreasing function of $\dot\Lam^2$.  Since $A(\Lam,\dot\Lam=0)>0$, a positive $\dot\Lam^2$ solution is guaranteed, so the window $\Lam<\Lam_m$ is classically allowed. When $\Lam>\Lam_m$, $A(\Lam,\dot\Lam^2)$ is a non-monotonous in $\dot\Lam^2$. The solution depends on the detail of the functions. For high T yet still below $T_c$, $\Lam_m>\Lam_h$, same analysis gives a classically allowed window $\Lam<\Lam_h$. There is numerical evidence that $\Lam_m$ and $\Lam_h$ move toward the boundary as temperature increases, which translates to the shrinking of the classically allowed window. 

One can solve the equation of motion for the domain wall (\ref{eq_dotlam}) analytically for an arbitrary constant energy $H= H_0$. One finds three branches of solutions 
\be\label{sol1}
\dot\Lam_k^2 = \frac{P- u_k^2}{Q}; \qquad u_k = -\frac13\le(v_k C - \frac{3a}{v_k C}\ri)\, ,
\ee
where 
\bea\lab{sol2} 
C &=& 3 \le(\frac{b}{2}\ri)^{\frac13}\le(1+\sqrt{ 1+\frac{4a^3}{27b^2} }\ri)^\frac13, \\
a&=& -\frac{QV-F-H_0}{M} -P, \quad b = \frac{Q}{M}\, . \no
\lab{sol3}
v_1 &=& 1, \,\, v_2 = e^{\frac{2\pi i}{3}}, \,\, v_3 = e^{-\frac{2\pi i}{3}}\, .
\eea 
The solution with our boundary condition is obtained by setting $H_0 = -F$ in the expression for $a$ above.  

Now we will demonstrate that near the boundary region $\Lam\to 0$ one obtains quantum tunneling with a highly suppressed tunneling probability for $T>T_c$,  and classical rolling for $T<T_c$, as expected: For small $\dot\Lam^2$, we can approximate
\begin{align}\label{A_app}
A(\Lam,\dot\Lam^2)=V(\Lam)-\frac{1}{\sqrt{P(\Lam)}}+\(M(\Lam)-\frac{Q(\Lam)}{2P(\Lam)^{3/2}}\)\dot\Lam^2.
\end{align}
The solution to $A(\Lam,\dot\Lam^2)=0$ is given by
\begin{align}\label{lamdot}
\dot\Lam^2=-\frac{V(\Lam)-\frac{1}{\sqrt{P(\Lam)}}}{M(\Lam)-\frac{Q(\Lam)}{2P(\Lam)^{3/2}}}.
\end{align}
In the region $\Lam\to0$, the numerator approaches $F$ which is negative(positive) for $T<T_c$($T>T_c$). On the other hand, the denominator tends to negative infinity: using near boundary asymptotic expansion of the metric functions in  \cite{GKLN} and one obtains as $\Lam\to 0$, 
\begin{eqnarray}\label{Mlim} 
M  \to  -3(\log\Lam)^2 (4E-3 ST) + \cdots  \\
\label{Qlim}
\frac{Q}{2P^{3/2}}  \to +\frac34(\log\Lam)^2 (4E-3 ST) + \cdots 
\end{eqnarray} 
where ellipsis denote terms that are constant in the limit. The combination in the brackets is nothing but the renormalized trace of the stress tensor of the blackhole, $\la T^\mu_\mu(T)\ra -\la T^\mu_\mu(0)\ra$ \cite{GKLN} hence it is positive definite. Therefore the denominator in (\ref{lamdot}) indeed goes to negative infinity as $-15/4(4E-3TS) (\log\Lam)^2$. 

\begin{figure}
\includegraphics[width=0.4\textwidth]{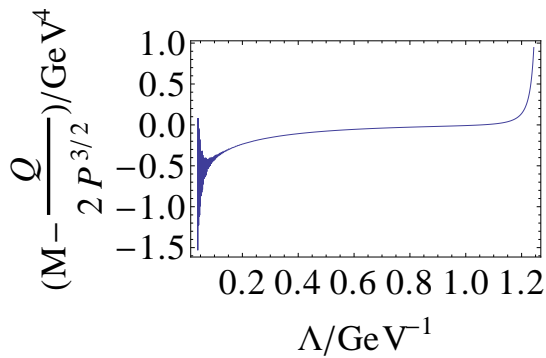}
\caption{\label{fig_Mc}$M-\frac{Q}{2P^{3/2}}$ versus $\Lam$ for $\af_h=0.7$, which corresponds to $T>T_c$. The function tends to negative value as $\Lam\to 0$.}
\end{figure}

The solution to (\ref{lamdot}) 
\begin{align}\lab{solbnd}
\dot\Lam^2\sim-\frac{F}{M-\frac{Q}{2P^{3/2}}},
\end{align}
 is indeed infinitesimal then, justifying \footnote{One does not need to assume small $\dot\Lam^2$ to carry out the analysis. Instead one can use the exact solution (\ref{sol1}), perform the near boundary expansion and obtain the same result.} the approximation \eqref{A_app}. With the solution, we have for the integrated action
\begin{align}\label{tunnel_prob}
i\int \frac{d\Lam d^3x}{\dot\Lam}\(\frac{S}{\bt\text{Vol}}-F\)=2\sqrt{F}\int d\Lam d^3x\sqrt{M-2\frac{Q}{2P^{3/2}}}.
\end{align}
We have subtracted the free energy $F$ of the blackhole from the action as it defines the zero point energy of the system.
Figure \ref{fig_Mc} shows a plot $M-\frac{Q}{2P^{3/2}}$ versus $\Lam$. The $\Lam\to0$ region ineed tend to a large negative value given above. Even though the integrated exponent (\ref{tunnel_prob}) in the limit $\Lam\to 0$ seems to converge as $\sqrt{15} \sqrt{(4E-3TS)/(TS-E)}(\Lam\log\Lam + const)$, this still provides a large suppression. By the same token, for $T<T_c$ the numerator in (\ref{solbnd}) becomes positive and one finds classical rolling.    
The logarithmic UV asymptotics in (\ref{Mlim}), (\ref{Qlim}) follow from the construction of the holographic QCD background by modeling asymptotic freedom of QCD in the UV. 

\begin{figure}
\includegraphics[width=0.4\textwidth]{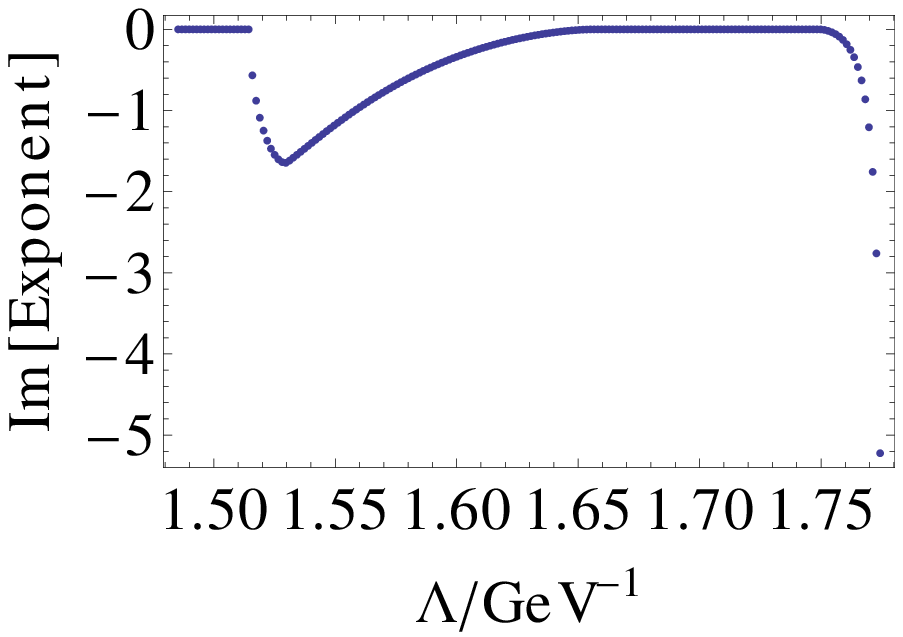}
\includegraphics[width=0.4\textwidth]{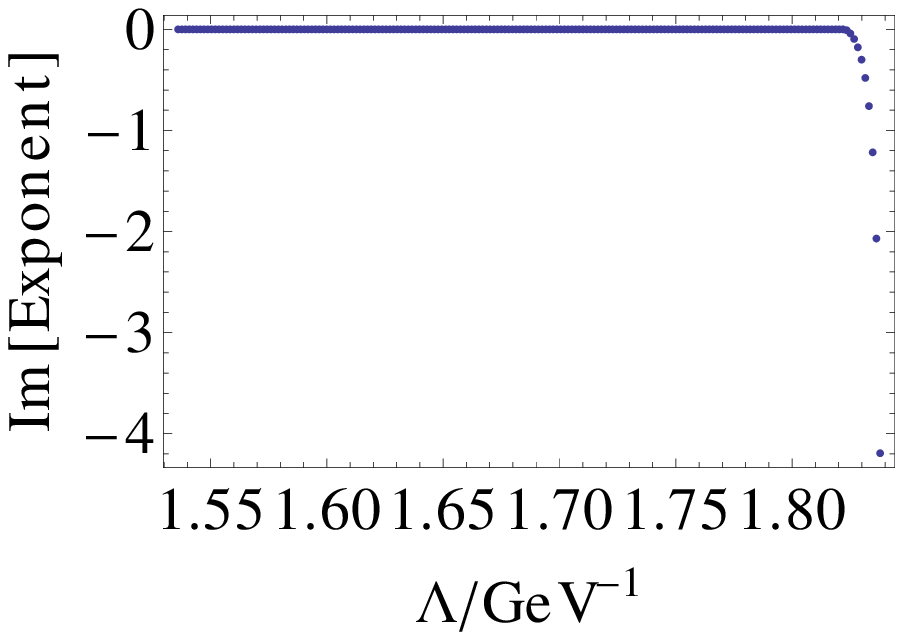}
\caption{\label{fig_qtw}The quantum tunneling windows for $\af_h=1.2$(top) and $\af_h=1.3$(bottom), both at $T<T_c$. The supercooled blackhole with higher temperature $\af_h=1.2$ has two tunneling windows, while the one with lower temperature $\af_h=1.3$ has only one tunneling window near the horizon. The lower temperature blackhole has higher tunneling probability.}
\end{figure}

We now return to our interest $T<T_c$, and study quantum tunneling near the horizon. For this we start with the analytic solution (\ref{sol1}) and perform near horizon expansion. For the three branches in (\ref{sol1}) one obtains the following behavior    
\be\lab{solhr} 
\dot\Lam^2_k \to f^2 - v_k^*\, p^{-\frac13}(ST)^{-\frac23} f^\frac73 + {\cal O}(f^\frac52)\, , 
\ee
where the constants $v_k$ are defined in (\ref{sol3}) and $p$ is a non-universal constant determined by the horizon value of $P$ in  (\ref{vmpq}) as $P\to p/f$. We should require presence of quantum tunneling near the horizon. One can show that the first solution  $\dot\Lam_1$ is real to all orders in $f$ hence never leads to quantum tunneling. Among the other two solutions one should choose the one that leads to a negative value for the imaginary part of the exponent
\begin{align}\label{exp_def}
\text{Exponent}=\frac{1}{\dot\Lam}\(\frac{S}{\bt\text{Vol}}-F\),
\end{align}
with the tunneling probability given by
\begin{align}\label{prob_def}
\text{Probability}=\exp\(-\int_0^{z_h} d\Lam d^3x\text{Im[Exponent]}\).
\end{align}
Substituting the asymptotic behavior (\ref{solhr}) in the action for $k=2,3$ and evaluating the imaginary part of the exponent above, one finds 
\be\label{exp}
\textrm{Im[Exponent]} \to \pm \frac{3\sqrt{3}}{8}(p^{2/3} (ST)^{4/3})^{-1} f^{-\frac13} + \cdots 
\ee 
Requiring a negative imaginary part, one chooses the branch $k=2$ in (\ref{sol1}). Note this also fixes the branch in the square root $\sqrt{\dot\Lam^2}$. Although the integrand is diverging at the horizon, see Fig. \ref{fig_qtw},  the integral is convergent, hence the tunneling barrier near the horizon is penetrable. 
One can also find the dependence of steepness of the barrier on $T$ from this expressions. Extracting the $T$ dependence from the constant $p$ (\ref{exp}) as $b(z_h) = S^{1/3}$ one finds that the steepness of the barrier grows with T as $s(T)^{2/3} T^{1/3}$ where $s\equiv S/T^3$ is a dimensionless entropy that itself grows with T, see e.g. \cite{GKLN}.  
Therefore the tunneling probability near horizon is much more suppressed in the high T regime. On the other hand $s$ vanishes as $T\to T_{min}$, leading to flattening of the barrier in the supercooled BH.  

\begin{figure}
\includegraphics[width=0.4\textwidth]{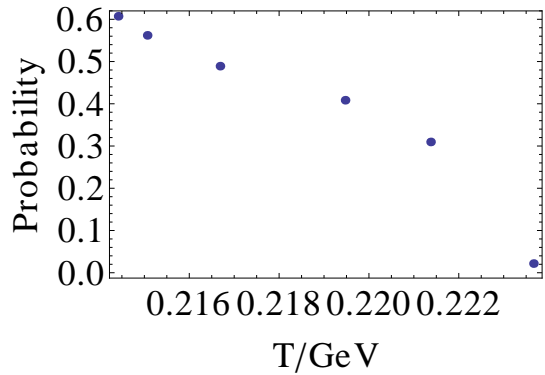}
\caption{\label{prob}The tunneling probability as a function of blackhole temperature with the critical temperature being at $T=0.273GeV$.}
\end{figure}

We now turn to numerical exploration of the results. The probability (\ref{prob_def}) only receives contribution from quantum tunneling windows. We find numerically that the quantum tunneling window broadens as temperature of BH increases. Figure \ref{fig_qtw} shows the quantum tunneling windows for $\af_h=1.3$ and $\af_h=1.2$. 
To fix the volume integral $\int d^3x$ in (\ref{prob_def}), we estimate the volume by a sphere with a radius of $4fm$, which is typical for a cluster size before particle emission according to HBT radii analysis \cite{hbt}. This gives a volume factor $\int d^3x\sim 3.35\times 10^4GeV^{-3}$.
We show in Figure.\ref{prob} the tunneling probability from the blackhole as a function of the temperature. The absolute value of the tunneling probability should be taken with caveat, as it is sensitive to the choice of volume factor. However the monotonous behavior as a function of temperature is robust.

\section{Summary}

In this paper, devoted to inhomogeneous out-of-equilibrium dynamical situations near the  deconfinement transition, we considered two distinct 
stages of the cooling process.

In the first ``beginning" stage we studied properties of the sound modes, in a supercooled  high-$T$ phase.   While one may expect at certain parameters appearance of sound instabilities,
we have not found  it happening, in the model under consideration. However we do observe rather remarkable changes in the behavior of the sound mode,
in the form of {\em stoppage of the propagation} and pure dissipative behavior, not known before in such setting.

In the second part we consider the end of the inhomogeneous phase, in which the last clusters of the  high-$T$ phase are expected to disappear.
We propose unusual configurations, in which the two phases are not separated by a boundary both in the usual and holographic coordinates -- denoted by $r$ and $z$ --
but in the holographic direction only. Two solutions of the equations of motion are thus separated by a ``membrane",
 whose motion we study. We suggest this qualitatively corresponds to the 
collapse of the QGP cluster under consideration.

Finally, we discuss quantum tunneling, using effective one-dimensional action derived in the previous subsection. 
 We find that classical rolling window exists for supercooled blackhole near the boundary, but doesn't exist for normal blackhole.
We map out tunneling windows as a function of the membrane position for supercooled blackhole at different temperatures. We find a universal tunneling window near the horizon, with an infinite but penetrable barrier. Other possible windows exist for high temperature supercooled blackhole. 
  \\ \\
{\bf Acknowledgements.}
ES would like to thank Jacob Sonnenschein for multiple discussions of the finite cluster problem, while SL acknowledges 
useful conversations with E. Kiritsis and T. Springer. SL also thank E. Megias and K. Veschigini for sharing mathematica notebook on blackhole thermodynamics.
The work of ES is partially supported by the U.S. Department of Energy under Contract No. DE-FG-88ER40388. The work of SL is supported by RIKEN Foreign Postdoctoral Researchers Program.



\end{document}